%% file: _New_Main.tex
\documentclass[conference,compsoc]{IEEEtran}
\ifCLASSOPTIONcompsoc
  \usepackage[nocompress]{cite}
\else
  \usepackage{cite}
\fi

\usepackage{url}
\usepackage[hidelinks]{hyperref}
\usepackage{xcolor}
\usepackage{multirow}
\usepackage{graphicx}
\usepackage{lipsum}
\usepackage{tabularx}
\usepackage{caption}
\usepackage{CJKutf8}
\usepackage{xspace}
\usepackage{booktabs}
\usepackage{afterpage}
\usepackage{tabulary,adjustbox}
\usepackage{soul}

\ifCLASSINFOpdf
\else
\fi

\newif\ifstatus
\statustrue

\begin{document}
\begin{CJK*}{UTF8}{gbsn}
%
\title{Privacy Perspectives and Practices of Chinese Smart Home Product Teams}

\author{\IEEEauthorblockN{Shijing He}
\IEEEauthorblockA{King's College London\\
London, United Kingdom\\
shijing.he@kcl.ac.uk}
\and
\IEEEauthorblockN{Yaxiong Lei}
\IEEEauthorblockA{University of St Andrews\\
St Andrews, United Kingdom\\
yl212@st-andrews.ac.uk}
\and
\IEEEauthorblockN{Xiao Zhan}
\IEEEauthorblockA{King's College London\\
London, United Kingdom\\
xiao.zhan@kcl.ac.uk}
\and
\IEEEauthorblockN{Chi Zhang}
\IEEEauthorblockA{University of Warwick\\
Coventry, United Kingdom\\
chi.zhang.9@warwick.ac.uk}
\and
\IEEEauthorblockN{Juan Ye}
\IEEEauthorblockA{University of St Andrews\\
St Andrews, United Kingdom\\
jy31@st-andrews.ac.uk}
\and
\IEEEauthorblockN{Ruba Abu-Salma}
\IEEEauthorblockA{King's College London\\
London, United Kingdom\\
ruba.abu-salma@kcl.ac.uk}
\and
\IEEEauthorblockN{Jose Such}
\IEEEauthorblockA{King's College London \& INGENIO \\
(CSIC-Universitat Politecnica de Valencia)\\
jose.such@csic.es}
}
\maketitle

\begin{abstract}
Previous research has explored the privacy needs and concerns of device owners, primary users, and different bystander groups with regard to smart home devices like security cameras and smart speakers, but little is known about the privacy views and practices of smart home product teams, particularly those in non-Western contexts. This paper presents findings from 27 semi-structured interviews with Chinese smart home product team members
. We examine their privacy perspectives, practices, and risk mitigation strategies. Our results show that participants emphasized compliance with Chinese data privacy laws, which typically prioritized national security over individual privacy rights. China-specific cultural, social, and legal factors also influenced participants' ethical considerations and attitudes toward balancing user privacy and security with convenience. 
Drawing on our findings, we propose a set of recommendations for smart home product teams, along with socio-technical and legal interventions to address smart home privacy issues---especially those belonging to at-risk groups---in Chinese multi-user smart homes.
\end{abstract}


%
\IEEEpeerreviewmaketitle

\input{content/1-intro}
\input{content/2-related}
\input{content/3-method}
\input{content/4-results}
\input{content/5-discussion}

\input{content/6-conclusion}


  \section*{Acknowledgments}

The authors would like to thank all participants from Chinese smart home product teams for taking part in this study, as well as the anonymous reviewers for their insightful feedback. We also extend our gratitude to Yixin Zou, Julia Bernd, and Adam Jenkins for their valuable comments on earlier drafts of this paper.



%



\bibliographystyle{plain}
\bibliography{reference}


\newpage 

\appendices 

\section{Meta-Review}

The following meta-review was prepared by the program committee for the 2026
IEEE Symposium on Security and Privacy (S\&P) as part of the review process as
detailed in the call for papers.

\subsection{Summary}
This paper conducts a semi-structured interview study with 27 with Chinese smart home product team members, including managers, engineers, designers, legal/privacy experts, and marketers. The paper focuses on teams' approaches and perceptions of privacy, and the challenges they face when addressing privacy. The authors find that participants emphasized compliance with Chinese data privacy laws and situate their findings within the Chinese context, including how ideas of Confucianism and filial piety play a role in how people think about and design for privacy.

\subsection{Scientific Contributions}
\begin{itemize}
\item Independent Confirmation of Important Results with Limited Prior Research
\item Provides a Valuable Step Forward in an Established Field
\end{itemize}

\subsection{Reasons for Acceptance}
\begin{enumerate}
\item This paper provides independent confirmation of important results with limited prior research by investigating how Chinese smart home product teams interact with and respond to privacy regulations.
\item This paper provides a valuable step forward in an established field by examining privacy in a non-Western cultural context, including Confucianism and filial piety influences on privacy and design practices.
\end{enumerate}

\subsection{Noteworthy Concerns} 
\begin{itemize}
    \item 
    The interview guide may be leading, with early emphasis on privacy and legal requirements potentially priming responses to later questions on roles and data practices.
\end{itemize}






\end{CJK*}
\end{document}

%% file: content/1-intro.tex
\section{Introduction}
Smart home devices have seen widespread global adoption, with projections estimating approximately 672.5 million users of smart home devices worldwide by 2027
~\cite{jeremiah2023smarthomedata}. This rapid adoption has significantly raised privacy concerns among different stakeholders, including 
device owners, primary users, secondary users, and bystanders in multi-user smart homes 
(e.g.,~\cite{yao2019defending,zeng2017end,mare2020smart,huang2020amazon,despres2024my,bernd2020bystanders}).
Despite this extensive literature from an \emph{end-user} perspective, few studies have considered 
smart home \emph{product teams}. 
This is a notable gap, as the perceptions, perspectives, and practices of those designing and developing smart home devices may play a crucial role in multi-user smart home privacy. 
Although some work has been conducted with smart home \emph{designers}~\cite{chalhoub2020innovation,chalhoub2024useful,albayaydh2024innovative,aljeraisy2021privacy}, 
there remains a 
gap in understanding how \emph{smart home product teams}---comprising software/hardware engineers, product managers, legal/regulatory compliance advisors, and marketing staff---address privacy-related challenges. 


Furthermore, the current body of research on smart home privacy is geographically focused on 
WEIRD (Western, Educated, Industrialized, Rich, Democratic) countries 
\cite{nurgalieva2023narrative,chalhoub2020innovation,chalhoub2024useful,seymour2023voice}. Smart home privacy in non-WEIRD countries has received limited scholarly attention~\cite{hasegawa2024weird}. As a non-WEIRD nation, China offers a distinct set of social, cultural, and legal characteristics that can surface novel privacy challenges and design considerations in the realm of smart home technologies~\cite{he2025exploring}, including:
i) the country's accelerated digital transformation~\cite{mckinsey}; ii) the high rate of smart home device adoption, with over 78 million Chinese households using such devices in 2023~\cite{xiaomi2023}; iii) the large number of smart home product teams in China~\cite{daniel2023chinasmarthome}; iv) the introduction of new privacy-focused laws and regulations within the country (\S\ref{subsec:regulations}); 
and v) China's tech landscape---driven by firms like Xiaomi instead of Western giants---which has given rise to alternative data governance practices and innovation pathways~\cite{zhang2024high}. 
Our study is the first to examine the privacy views and practices of Chinese smart home product team members, focusing on those involved in designing, developing, or evaluating smart home devices. These teams typically include software/hardware engineers, UX designers, product/project managers, legal/privacy experts, and marketing or operations specialists. 
Our research questions (RQs) are as follows: 


\begin{enumerate}
    \item[\textbf{RQ1.}] \emph{How do Chinese smart home product teams perceive and approach 
    multi-user smart home privacy throughout the design and development life-cycle?}
    \item[\textbf{RQ2.}] \emph{What challenges do Chinese smart home product teams encounter when addressing the privacy needs and concerns of diverse stakeholders in multi-user smart homes?}
    \item[\textbf{RQ3.}] \emph{How do members of Chinese smart home product teams respond to the privacy needs and concerns of diverse stakeholders in multi-user smart homes, considering technical, social, and legal dimensions?}
\end{enumerate}
To address our research questions, we conducted semi-structured interviews with 27 members of Chinese smart home product teams from companies of varying types (state-owned, multinational, private) and sizes (small, medium, large), all involved in designing and developing a wide range of smart home devices. 
Our key findings are as follows:

\textbf{RQ1.} Participants highlighted the importance of legal/ regulatory compliance in the design and development of smart home devices, often participating in routine reviews and showing a strong ethical commitment to user privacy. However, many---particularly UX designers---lacked adequate knowledge of privacy laws/data practices and formal training. They also noted that China's unique sociocultural context (e.g., collectivism and Confucian values) contributed to lower privacy concern among Chinese smart home device users and limited legal literacy among product teams, shaping the design of privacy features.

\textbf{RQ2.} Participants encountered several challenges, including: (1) managing complex and dynamic relationships between government authorities, companies, and users -- for example, situations where law enforcement agencies could access user data, bypassing internal privacy safeguards within companies; (2) limited attention to the privacy needs of at-risk groups (e.g., older adults, domestic workers), stemming from difficulties in conducting empirical research with these populations and ethical limitations; and (3) the need to balance users' privacy with UX, such as ease of use and convenience, particularly within the shared environment of multi-user smart homes. 

\textbf{RQ3.} To address the challenges identified above, participants used various privacy practices and risk mitigation strategies. These included \emph{technical} measures such as restricting data access and implementing decentralized data processing. They also advocated for \emph{social} initiatives to improve user privacy awareness (e.g., community-level outreach activities) and called for clearer \emph{legal} guidelines for smart home product design/development. Although participants acknowledged recent improvements in local privacy legislation, they highlighted a gap between legal frameworks and actual development practices, often caused by the fast-paced evolution of smart home technologies in China. 



\textbf{Contributions.}
We conducted the first exploratory interview study focusing on Chinese smart home product team members to investigate their perspectives and practices related to privacy. 
Unlike prior studies, our participant sample includes smart home product team members in various roles mainly from large companies, providing a more comprehensive overview of their privacy perspectives and practices (see Table \ref{tab:demographic}). Additionally, this study is the first to explore how these product team members navigate and implement privacy laws and ensure legal compliance during smart home product design and development. Our findings provide valuable and rich empirical insights into privacy practices in the Chinese smart home industry, with implications for product teams, researchers, and policy makers, especially in light of China's distinct sociocultural environment. 
Based on our empirical evidence, we 
extend the discussion on Privacy by Design by emphasizing the need for responsible and transparent guidelines adapted to China's unique sociocultural context---such as Confucian values, collectivist norms, and China's culturally specific understanding of ``privacy'' (\S\ref{chinesecontext}). 
Further, our findings highlight disparities and power dynamics between large companies---which typically have greater resources, dedicated compliance teams, and structured internal legal support---and small- and medium-sized enterprises (SMEs), which often face significant challenges in privacy compliance due to resource constraints and limited legal expertise. Based on our empirical evidence, we strongly recommend clearer legal frameworks to guide privacy practices, specifically designed to support compliance efforts in SMEs. 
Finally, we call for enhanced awareness of privacy issues among both the general public and product team members through educational initiatives in communities and schools, as well as targeted training for legal/regulatory compliance and data management (\S\ref{privacyindevelopement}). 

%% file: content/2-related.tex
\section{Background and Related Work} 
\label{related}
\subsection{Multi-User Smart Home Privacy} \label{multiuser}

Previous studies have explored the privacy, security, and safety needs, preferences, and concerns of smart home device owners and primary users~\cite{abdi2019more,abdi2021privacy,yao2019defending,thakkar2022would,zeng2017end,tabassum2019don,lau2018alexa,huang2020amazon,mare2020smart}, including Chinese users~\cite{li2023will,huang2019perception,he2025living}, 
emphasizing the importance of data ownership by users, data practice transparency, and informed user consent. The studies have also highlighted how social norms, including privacy norms, along with factors such as power dynamics, social structures, demographics, and technical expertise, influence views on smart home device adoption and use~\cite{abdi2021privacy,apthorpe2018discovering,tabassum2019don,slupska2022they}. 
Other papers have explored the privacy implications of smart home devices for different bystander groups who share a living space, including 
visiting, live-in, and uninvolved bystanders~\cite{saqib2025bystander}, highlighting privacy and security concerns of at-risk groups such as 
domestic workers~\cite{bernd2022balancing,bernd2020bystanders,slupska2022they,despres2024my,abu2025they,he2025exploring} and household members living in hostile environments~\cite{alshehri2020smart,leitao2019anticipating}. 
Additionally, some studies engaged device owners/users and bystanders in the co-design and evaluation of prototypes for select design interventions proposed in the academic literature~\cite{mare2020smart,h2022monitoring,windl2023investigating,thakkar2022would,marky2022you,lau2018alexa,ahmad2020tangible,pierce2020sensor,yao2019privacy}.

Building on the findings of the aforementioned studies involving \emph{primarily} device owners and bystanders, researchers proposed \emph{technical} solutions to improve multi-user smart home privacy, including minimizing data collection for essential functions~\cite{mare2020smart}; selectively limiting data collection in sensitive situations~\cite{cobb2021would,h2022monitoring}; designing privacy nudges and notifications~\cite{geeng2019s,lau2018alexa,zeng2017end}; implementing temporary data storage mechanisms~\cite{h2022monitoring,cobb2021would}; 
using smart home dashboards for data visualization~\cite{mare2020smart,windl2023investigating,thakkar2022would}; creating separate user profiles (e.g., guest profiles with limited data access and control)~\cite{alshehri2020smart,geeng2019s,huang2020amazon}; employing tangible practices (e.g., camera lens caps)~\cite{ahmad2020tangible,ahmad2022tangible}; and implementing communication and negotiation mechanisms to state privacy preferences~\cite{h2022monitoring,albayaydh2024co,zhou2024bring}.
Other studies proposed \emph{social} interventions to address privacy concerns in multi-user smart homes, e.g., design guidelines aimed at supporting smart home product teams~\cite{bernd2022balancing,chalhoub2020innovation,albayaydh2024co}. 


\subsection{Privacy in Technology Design/Development} \label{privacydevelopment}
Previous research on privacy in digital technology design and development focused primarily on the challenges encountered by \emph{software developers of mobile apps} in eliciting and incorporating privacy-related requirements. These challenges include privacy- and security-related issues in development~\cite{li2018coconut,tahaei2019survey,kaur2025threat,schmuser2024m,acar2016you}, the failure to comply with privacy laws and regulations \cite{alomar2022developers,tahaei2022charting}, and guidance documentation that inadvertently leads developers to privacy-compromising decisions~\cite{tahaei2021developers,klemmer2023make}. In addition, privacy concerns have been identified in 
APIs and tools used by programmers~\cite{chowdhury2021developers,gutfleisch2022does}. Research also showed that developers often deprioritize privacy due to stakeholder pressures~\cite{ekambaranathan2020understanding,klemmer2023make}, placing greater emphasis on revenue generation and functionality than on privacy protection ~\cite{tahaei2022embedding,kekulluouglu2023we,tahaei2023stuck}. (Refer to~\cite{nurgalieva2023narrative} for a review of the factors influencing software developers' security and privacy practices.) Unlike prior studies, our paper investigates the privacy perceptions and practices of Chinese smart home product teams, thereby extending the focus from software developers to a broader range of product team members in diverse roles and organizations, focusing on smart home devices. 

In smart homes, Chalhoub et al. examined how product teams working in UK companies integrate UX principles into the security and privacy design of \emph{smart home cameras}, proposing solutions such as customizable privacy settings and transparent user consent mechanisms~\cite{chalhoub2020innovation,chalhoub2024useful}. 
These studies were conducted in a WEIRD context, focusing exclusively on a single type of smart home device---smart cameras---and on \emph{end-user experience}. They did not study how product teams perceived the privacy concerns of other stakeholders in smart homes, including bystanders, and did not take into account the power dynamics that exist among these groups. 
Zhang and Xie examined the main activities of and challenges faced by IoT developers---mainly hardware engineers, embedded software developers, and app developers---in Chinese (non-WEIRD) start-ups~\cite{zhang2018toward}. Although the study did not focus on security and privacy in the smart home context, and recruited participants only from \emph{startups}, it found that 
participants frequently overlooked security issues due to limited resources and skills. 

Finally, recent work~\cite{albayaydh2023examining,albayaydh2024co} showed how socioeconomic factors, social norms, and religious beliefs influence power dynamics in Jordanian (non-WEIRD) smart homes. However, these studies did not explore the privacy perceptions and practices of local smart home product teams, focusing instead on end-users and a bystander group: domestic workers. While follow-up work~\cite{albayaydh2024innovative} focused on \emph{smart home designers} and their lack of privacy-support tools and guidelines, our study offers novel insights into \emph{Chinese product teams'} perspectives and practices, examining privacy considerations across multiple roles. Additionally, our research highlights how cultural values (e.g., filial piety), priorities (e.g., prioritizing legal compliance over individual privacy), and hierarchical, collectivist dynamics influence privacy trade-offs within these teams.

\subsection{Confucianism and Privacy in China} \label{subsec:Confucianism}
Confucianism, a foundational philosophy in Chinese culture, emphasizes social harmony, hierarchical relationships, and the importance of family and community~\cite{yao2000introduction}. Central to Confucian thought are concepts like \emph{filial piety} (``xiao''), which underscores respect and duty toward one's parents and elderly people, and \emph{collectivism}, which prioritizes group needs over individual desires~\cite{ho1996filial}. Some studies have highlighted that traditional Confucian thinking, which significantly affects the understanding of privacy within the Chinese family unit, is often associated with the indigenous Chinese concept of ``yin si'' with notions of unspeakable immorality, indecency, or shame, providing minimal support for the development of legal norms that favor individual privacy protection~\cite{naftali2010caged,wang2016examining,yao2005privacy}. 
Consequently, privacy norms in China have developed at a slower pace compared to Western societies, where privacy is deeply rooted in concepts of liberal individualism~\cite{bennett2011defense} and central to liberal political theory~\cite{hefferman1995privacy}.

\subsection{Legal and Regulatory Frameworks in China}
\label{subsec:regulations}
In China, companies' privacy practices are primarily governed by the Personal Information Protection Law (PIPL)\footnote{\href{http://en.npc.gov.cn.cdurl.cn/2021-12/29/c_694559.htm}{Personal Information Protection Law (PIPL)}}, Cybersecurity Law (CSL)\footnote{\href{http://www.npc.gov.cn/zgrdw/npc/xinwen/2016-11/07/content_2001605.htm}{Cybersecurity Law (CSL) of the People's Republic of China}}, Data Security Law (DSL)\footnote{\href{https://www.gov.cn/xinwen/2021-06/11/content_5616919.htm}{Data Security Law (DSL) of the People's Republic of China}}, and Provisions on Online Protection of Children's Personal Information (POPCPI)\footnote{\href{https://www.gov.cn/xinwen/2019-08/23/content_5423865.htm}{Provisions on Online Protection of Children's Personal Information}}. These Chinese legal frameworks establish data privacy protections and define the scope of state oversight over data processors and companies~\cite{junck2021china,xixin2022bundle}. While GDPR, CCPA, and PIPL share common goals of protecting personal data and individual rights, they differ in their enforcement mechanisms, the extent of government access, and overall emphasis. Although PIPL is influenced by GDPR and incorporates similar individual rights, it places greater emphasis on national security and allows broader government access to data~\cite{calzada2022citizens, gao2022attractive, wang2024justifying}. 
CSL and DSL contain provisions related to personal information protection 
but they both prioritize national security over individual privacy rights~\cite{calzada2022citizens}. In contrast, PIPL focuses on individuals' privacy within the Chinese context (e.g., Chapter 3). However, certain terms in PIPL---particularly those related to separate consent and the criteria for lawful anonymization---are open to interpretation, creating uncertainty in their application~\cite{yao2023overcoming,wang2024justifying}. 
PIPL supports individual data protection, but allows government access under certain circumstances (e.g., Chapter 2). Furthermore, enforcing data protection laws in China is complex due to the involvement of multiple administrative authorities, overlapping responsibilities, inconsistent enforcement practices, and internal contradictions within the legal framework~\cite{creemers2022china,zhang2024china}. 

%% file: content/3-method.tex
\input{appendix/demograpic}

\section{Methodology} \label{method} 
We conducted semi-structured interviews with 27 Chinese smart home product team members from mainland China, using Mandarin for all interviews\footnote{The translated screening survey, interview script, and codebook, as well as a comparative table of PIPL, DSL, CSL, and POPCPI, are available in \href{https://osf.io/5x6pz/overview}{\textbf{this OSF repository}}.}. 
This section outlines our recruitment approach, participant demographics, interview protocol, data analysis details, ethical considerations, and study limitations. 

\subsection{Recruitment and Participants}
We aimed to recruit a diverse sample of 
Chinese nationals who were 18 years or older and were employed as product team members in smart home companies based in mainland China, ensuring variation in roles, company sizes, and experiences. 
%
We 
identified participants with these job titles using professional platforms (Maimai, LinkedIn)
, our industry connections with Chinese smart home companies
, and snowball sampling, in which participants referred peers in different roles, either within their own company or from other organizations
~\cite{parker2019snowball}.
We asked all interested individuals to complete a screening survey in Chinese. 
Based on industry standards and expected time commitment, we compensated each participant £50. 
\label{tab:participants}
Our 27 participants, as detailed in Table~\ref{tab:demographic},  
included 11 women and 16 men; and 
9 had a bachelor's degree, 17 had a master's degree\footnote{In China's IT sector, it is common for product team members to hold a master's degree~\cite{masterdegree}.}, and 1 had a PhD. 
Participant roles included: 8 product/project managers (PM), 7 software/algorithm/hardware engineers (EN), 5 UX researchers/designers (UX), 4 legal privacy (data protection) officers/privacy engineers (LE), and 3 marketers/operations specialists (MS). Although we recruited some participants with company-specific job titles, these titles could be broadly categorized. For example, ``algorithm engineer'' falls under the broader category of ``software engineer.'' The responses from participants with such titles were consistent with the overall insights provided by other participants in the same broader category and did not influence the final themes. %
Unlike previous studies that have focused on non-WEIRD technologies (mainly software systems) and participants who worked primarily in \emph{start-ups}~\cite{kekulluouglu2023we,zhang2018toward,ribak2019translating}, most of our participants (20) worked in large companies with more than 500 employees. Furthermore, 3 participants worked in medium-sized companies (101–500 employees), and 4 worked in small companies with fewer than 100 employees. We included participants from small to large companies to capture a broad spectrum of perspectives (\S\ref{contributions}). Regarding company type, 8 participants were employed in state-owned enterprises (SOEs), 4 in multinational corporations (MNCs), and 15 in private enterprises (PEs). 22 participants were employed by different companies, while 5 worked in two firms: 2 in one (P9, P24) and 3 in another (P11, P13, P23). In addition, 4 employed by MNCs worked on devices for the Chinese market and operated under local compliance requirements. 

Our participants were involved in designing and developing various smart home devices, including smart speakers (8), smart home cameras (16), door locks/doorbells (9), smart health devices/monitors (6), smart thermostats (8), baby cameras/monitors (5), smart light bulbs (6), smart TVs (4), smart plugs (7), kitchen appliances (7), smoke detectors/alarms (4), and smart home robots (3). Notably, only P2-PM and P27-MS expressed distrust regarding the ownership of smart home devices, while the remaining 25 participants owned such devices in their own households.

\subsection{Interview Procedure} \label{interview_procedure}
Authors 1-3 are native Mandarin speakers fluent in English. Author 1 conducted all interviews remotely in Mandarin via VooV Meeting, while Authors 2 and 3 attended to take notes. The interviews had an average duration of 85.3 mins, ranging from 44.5 to 112 mins. 
We audio-recorded and transcribed all sessions
, and then we independently and manually checked the accuracy of all transcripts. 
The interviews began with participants outlining their roles within a product team, detailing the types of smart home devices they designed or developed, and their responsibilities. We then asked questions about participants' privacy and/or legal obligations. Although these questions were introduced early in the guide---after asking about roles, team structure, and responsibilities to build rapport---they were intended to examine participants’ roles in data governance. The questions were broadly framed to avoid assumptions about participants' views and were followed by exploratory probes. We then investigated participants' practices regarding the collection of user requirements, with a focus on UX and privacy-related requirements. We aimed to understand how participants examined the needs and privacy concerns of different stakeholder groups, including at-risk groups (e.g., children, domestic workers) in multi-user smart homes, drawing on previous work such as~\cite{sannon2022privacy,he2025exploring,bernd2020bystanders,leitao2019anticipating,chalhoub2020innovation}. Further, we examined how participants incorporated UX and privacy considerations throughout their design and development processes. It is worth noting that questions on UX requirements gathering (including privacy concerns) were directed primarily at participants in roles responsible for this task (e.g., product managers), while other participants (e.g., compliance officers) were asked about deriving requirements from laws, regulations, or sector-specific guidelines.

To reduce social desirability bias, we began with neutral, open-ended questions (e.g., ``What are the main challenges that you and your product team face when designing and/or developing smart home devices?'') and emphasized that there were no right or wrong answers. 
We also asked about data handling practices and legal compliance measures. To elicit deeper insights, we presented relevant examples from China (e.g., Didi case\footnote{\href{https://www.washingtonpost.com/world/2022/07/21/china-didi-fine-data-security/}{The Cyberspace Administration of China (CAC) fined Didi \$1.2 billion for the illegal collection and use of users’ personal data, in violation of laws on national data security and personal information protection.}}) to gauge their perspectives on technical interventions such as data collection, transmission, and deletion, as well as legal interventions like compliance training. These real-world examples encouraged participants to reflect on and propose technical, social, or legal interventions. We also sought their opinions on existing privacy laws and regulations in China (cf. \S\ref{subsec:regulations}). 

\subsection{Pilot Study} 

We carried out three pilots to collect feedback. The pilots are not included in the final analysis. 
The feedback we obtained helped us revise some questions related to design and development aspects such as requirements gathering (e.g., asking about user involvement), collaboration (e.g., asking about collaborating with legal compliance and marketing team members), and adherence to legal regulations (e.g., probing into details about participants' understanding of and compliance with PIPL). We also reflected critically on how the phrasing, sequencing, and specificity of questions might influence participants' responses. For example, we reworded compound or overly narrow questions to ensure participants in various roles could respond meaningfully.

\subsection{Data Analysis} Sample size in qualitative research remains a topic of debate~\cite{braun2006using}, particularly given the inherently subjective nature of determining data saturation---commonly defined as the point at which no new data is found~\cite{braun2021saturate}. In our study, we achieved data saturation after 25 interviews and conducted two additional interviews to confirm this. Specifically, three authors reviewed the data after every few interviews to assess whether saturation had been reached or if further interviews were needed, noting that insights began to repeat and that additional interviews did not provide new or significant variations to the existing data~\cite{guest2006many}. Our sample size also aligns with widely accepted qualitative research standards, which prioritize rich, contextualized understanding over statistical generalizability~\cite{vasileiou2018characterising}.

We inductively analyzed our interview transcripts using in-vivo coding to remain close to participants’ perspectives. Authors 1–3 coded and analyzed all transcripts in Chinese to preserve meaning and prevent potential loss that could have occurred if the transcripts had been translated into English first. Specifically, Authors 1–3 independently coded three randomly selected transcripts to develop an initial codebook. The three authors then met and, through joint discussions, identified codes, refined overlapping codes, clarified boundaries (e.g., distinguishing ``Prioritize privacy \& security requirements'' from ``Data management mechanisms''), and established rules for the consistent application of codes throughout the coding process. After resolving disagreements, they merged their individual codebooks into a single codebook. Using the merged codebook, they randomly selected and independently coded an additional transcript (different from the previous three), updating the codebook as needed. This process was repeated two more times to finalize the codebook and achieve code saturation.

After finalizing the codebook, each transcript was independently coded by two researchers (Interviews 1–13 by Authors 1 and 2, and 14–27 by Authors 1 and 3) to enhance rigor and reliability, prevent individual researcher bias, capture nuanced interpretations, and ensure consistent application of the codebook~\cite{mcdonald2019reliability}. We measured inter-rater reliability and checked agreement rates, resulting in an average Cohen's kappa of 0.88 for Authors 1 and 2 and 0.92 for Authors 1 and 3, indicating excellent agreement~\cite{fleiss2013statistical}.

As a research team, we then iteratively identified, developed, and refined themes until they addressed our research questions, aligning with the 20 themes presented in \S\ref{results}. In particular, we organized codes into themes specific to our research questions and reviewed the corresponding excerpts for each topic to further refine the themes. Finally, Author 1 translated the codebook and participant quotes reported in \S\ref{results} into English. These translations were subsequently reviewed and verified by Authors 2 and 3 to ensure accuracy and minimize potential translation gaps~\cite{xian2008lost}.

\subsection{Research Ethics} The Research Ethics Committee of King’s College London reviewed and approved this study (ID: LRS/DP-22/23-35814). While company names were collected during screening to assess the size and type of organizations, they were excluded from our reported findings to protect the privacy of both participants and their companies. Personal contact information (e.g., WeChat IDs) was used solely to communicate with potential participants during recruitment and was securely deleted once recruitment was completed. Following the transcription of all interview recordings, all transcripts were anonymized by removing identifying details, including names, company affiliations, and project-specific information. 
To safeguard participant confidentiality, only Author 1 retained the contact details of participants.

\subsection{Author Positionality} \label{author_positionality}
Since privacy can vary significantly based on cultural, social, and individual factors, we did not provide participants with a specific definition of privacy, as existing definitions in the academic literature are predominantly shaped by WEIRD values. Instead, we explored participants' privacy perspectives and practices as influenced by their backgrounds and experiences, given that our study focuses on a non-WEIRD country. Although participants did not explicitly mention the term ``Confucianism'' during the interviews, they discussed values aligned with it, such as filial piety (\S\ref{power dynamic}). Recognizing these culturally ingrained themes, we incorporated Confucianism into our analysis and reporting of results. Rather than treating it as a discrete code in our codebook, Confucianism emerged as an overarching theme during the process of developing and refining our themes.

In addition, our positionality and interpretation of the results were shaped by our backgrounds and experiences. All authors are expert researchers in human-centered computing, with extensive experience working with both majority populations and at-risk groups. Although all authors are trained researchers based primarily in Western institutions, the interviewers and data analysts contributed additional perspectives rooted in their personal backgrounds, having been born and raised in China. This dual perspective allowed them to approach the interviews and interpret the data with both academic rigor and deep cultural understanding. Also, Author 4 specializes in technology privacy law and public policy in both Chinese and Western contexts, including the UK and the US.

\subsection{Limitations} \label{limitations}
Given the qualitative nature of our study, the findings are not statistically generalizable beyond our sample. However, our sample of participants spans a wide range of company types/sizes and roles within product teams in China's smart home industry. This diversity enabled a rich and nuanced understanding of participants' experiences, perspectives, and practices~\cite{soden2024evaluating}. We also documented our study procedures in detail to support replication in other cultural and national contexts. 
Although most of our participants (24) were between 25 and 34 years old, this reflects broader demographic trends in China's tech sector, where companies often favor younger employees for product roles and view those over 35 years of age as less efficient~\cite{liu2023role}. Furthermore, we recognize that the interview guide might have included some leading questions, which could have affected participants' responses, despite our efforts to reduce bias by using open-ended and role-specific phrasing (cf. \S\ref{interview_procedure}). Lastly, as with all self-reported data, our findings might be influenced by social desirability bias~\cite{nederhof1985methods}. 

%% file: appendix/demograpic.tex
\renewcommand{\arraystretch}{1}
\begin{table*}[!ht]
    \centering
    \fontsize{6pt}{6pt}\selectfont
    \resizebox{\textwidth}{!}{%
    \begin{tabular}{p{0.8cm} p{0.5cm} p{0.5cm} p{0.8cm} p{0.8cm} p{0.8cm} p{2.6cm} p{0.9cm} p{0.3cm} p{6cm}}
    \toprule
        ~ & \textbf{Age} & \textbf{Gender} & \textbf{Education} & \textbf{Experience} & \textbf{Location} & \textbf{Role} & \textbf{Firm size} & \textbf{Type} & \textbf{Designed/Developed device(s)} \\ 
    \midrule
        P1-EN & 25-34 & Man & Master's & 6-10 yrs & Shanghai & Hardware engineer & 1000+ & MNC & Smart health devices/monitors\\
    
        P2-PM & 25-34 & Man & Master's & 1-5 yrs & Shanghai & Product manager & 1000+ & MNC & Smart speakers, Smart home cameras\\
    
        P3-PM & 25-34 & Man & Bachelor's & 6-10 yrs & Shanghai & Product manager & 11-50 & PE & Baby cameras/monitors, Smart lighting\\
      
        P4-PM & 25-34 & Man & Bachelor's & 1-5 yrs & Guangdong & Product manager & 1000+ & PE & Baby cameras/monitors, Smart home cameras, Door locks/doorbells\\
     
        P5-UX & 25-34 & Man & Bachelor's & 10+ yrs & Shanghai & UX design manager & 1000+ & PE & Smart speakers, Smart home cameras, Door locks/doorbells, Smart thermostats, Smart TVs, Smart lighting, Smart plugs \\
     
        P6-UX & 25-34 & Woman & Master's & 6-10 yrs & Sichuan & UX designer & 1000+ & SOE & Smart speakers, Smart TVs\\
     
        P7-PM & 25-34 & Man & Master's & 1-5 yrs & Guangdong & Product manager & 501-1000 & PE & Smart speakers, Door locks/doorbells\\
     
        P8-UX & 25-34 & Woman & Master's & 6-10 yrs & Shanghai & UX designer & 1000+ & MNC & Smart health devices/monitors, Smart thermostats\\
     
        P9-UX & 25-34 & Man & Master's & 1-5 yrs & Chongqing & UX designer & 1000+ & SOE & Smart speakers, Smart home cameras, Smart thermostats\\
     
        P10-PM & 35-44 & Man & Bachelor's & 10+ yrs & Jiangsu & Product manager/director & 1000+ & PE & Smart speakers, Smart home cameras, Door locks/doorbells, Smart health devices/monitors, Smart thermostats, Baby cameras/monitors, Smart lighting, Smart plugs, Smoke detectors/alarms\\
     
        P11-PM & 25-34 & Man & Master's & 6-10 yrs & Zhejiang & Project manager & 1000+ & SOE & Smart home cameras\\
     
        P12-PM & 25-34 & Woman & Bachelor's & 1-5 yrs & Beijing & Product manager & 1000+ & PE & Smart home cameras, Door locks/doorbells, Baby cameras/monitors\\
     
        P13-UX & 25-34 & Man & Master's & 6-10 yrs & Zhejiang & UX designer & 1000+ & SOE & Smart home cameras, Door locks/doorbells, Smoke detectors/alarms\\
     
        P14-EN & 25-34 & Man & Bachelor's & 6-10 yrs & Guangdong & Sr. software engineer & 51-100 & PE & Smart speakers, Smart home cameras, Door locks/doorbells, Smart health devices/monitors, Smart Thermostats, Smart lighting, Smart plugs, Smoke detectors/alarms, Kitchen appliances\\
    
        P15-EN & 35-44 & Man & Master's & 10+ yrs & Shannxi & Software director & 1000+ & SOE & Smart health devices/monitors, Smart TVs, Kitchen appliances\\
     
        P16-PM & 25-34 & Woman & Master's & 1-5 yrs & Guangdong & Product manager & 0-10 & PE & Smart speakers, Smart home robots\\

        P17-EN & 25-34 & Man & Doctoral & 1-5 yrs & Beijing & Algorithm engineer & 501-1000 & PE & Smart watches, Smart home cameras\\

        P18-LE & 25-34 & Man & Master's & 1-5 yrs & Beijing & Safety compliance engineer & 1000+ & PE & Smart TVs, Smart Thermostats, Smart lighting, Smart speakers, Smart plugs, Smart home cameras, Door locks/doorbells, Smart health devices/monitors, Smoke detectors/alarms\\

        P19-LE & 25-34 & Woman & Master's & 1-5 yrs & Shanghai & Compliance officer & 1000+ & PE & Smart home robots\\

        P20-LE & 25-34 & Man & Bachelor's & 6-10 yrs & Guangdong & Product compliance engineer & 101-500 & PE & Smart plugs, Smart home cameras, Door locks/doorbells, Kitchen appliances\\

        P21-EN & 25-34 & Man & Master's & 6-10 yrs & Guangdong & Software engineer & 1000+ & SOE & Smart home cameras\\

        P22-EN & 25-34 & Woman & Bachelor's & 1-5 yrs & Jiangsu & Software engineer & 1000+ & PE & Kitchen appliances, Smart home robots\\

        P23-LE & 25-34 & Woman & Master's & 1-5 yrs & Zhejiang & Compliance officer & 1000+ & SOE & Smart speakers, Smart home cameras, Baby cameras/monitors, Kitchen appliances\\

        P24-EN & 35-44 & Woman & Bachelor's & 10+ yrs & Chongqing & Software director & 1000+ & SOE & Smart thermostats, Smart speakers, Smart home cameras\\

        P25-MS & 25-34 & Woman & Master's & 6-10 yrs & Shanghai & Marketing executive & 101-500 & MNC & Smart plugs, Kitchen appliances\\

        P26-MS & 25-34 & Woman & Master's & 6-10 yrs & Shanghai & Digital marketing specialist & 51-100 & PE & Smart home cameras\\

        P27-MS & 25-34 & Woman & Master's & 1-5 yrs & Guangdong & Customer operations specialist & 101-500 & PE & Smart thermostats, Smart lighting, Smart plugs, Kitchen appliances\\
    \bottomrule 
    \end{tabular}}
    \caption{Participant demographics and the types of smart home devices they had designed/developed. 
    }
    \label{tab:demographic}
\end{table*}

%% file: content/4-results.tex
\section{Results} \label{results}
In this section, we present the privacy perspectives and practices of participants, with a focus on multi-user smart home privacy in \S\ref{tab:views and practices} \textbf{(RQ1)}; the challenges faced by participants during the development life-cycle, as well as their views on power dynamics and privacy conflicts in smart homes in \S\ref{challenges} \textbf{(RQ2)}; and \S\ref{tab:mitigating} describes the practices and strategies used by participants to address or mitigate privacy concerns of different stakeholder groups \textbf{(RQ3)}. Because this research is qualitative, we did not attempt to quantify how many participants expressed a given view or theme; our research goal was to capture the richness and nuance of different experiences. We use terms like ``most'', ``many'', ``some'', ``several'', and ``a few'' only to give a rough insight into prevalence. Across the sections, we treat privacy as contextual and selectively implemented, with protections shaped by legal mandates (e.g., POPCPI) and by cultural or dignity-based values (e.g., respectful blurring in elder care).

\subsection{Perspectives, Experiences, and Practices}
\label{tab:views and practices}

\subsubsection{
Privacy \& security considerations during development \emph{(product teams)}} \label{ProductTeams}
We describe how participants and their teams gathered requirements, considered privacy and security, and complied with relevant laws and regulations. 

\textbf{Gathering security and privacy requirements.} 
When we asked participants how they gathered user security and privacy-related requirements, most equated this process with complying with legal frameworks. These requirements were often tailored to the specific type of device being developed. For example, P5-UX highlighted the link between privacy and device type: \textit{``If the device is a surveillance camera or clearly collects user data, we will prioritize privacy and security.''} Many participants regularly attended product team meetings centered on user privacy, particularly those in product management or legal compliance positions. For example, P12-PM underscored the necessity of complying with privacy laws, explaining: \textit{``I would definitely place security- and privacy-related requirements first. If a feature did not comply with privacy laws, I would ask my product team to redesign and launch it immediately.''} P18-LE emphasized the importance of interpreting privacy-related regulations within the product team and described it as part of their responsibility: \textit{``Our team offers developers specialized legal and regulatory support by examining both domestic and international certification regulations and standards, providing comprehensive guidance to assist the development team with design and implementation.''}

Additionally, some participants described applying protections selectively and contextually, guided by both legal mandates and considerations of dignity. For example, P3-PM highlighted that their product team thoughtfully balanced children's privacy needs with parents' caregiving requirements while designing and developing smart desk lamp devices equipped with camera sensors. P3-PM shared,
\textit{``We gathered requirements from parents, and while most wanted constant monitoring features, we prioritized protecting children's privacy over providing extensive monitoring capabilities to parents. 
Even video calls require children's approval on their end.''} Participants explicitly connected these choices to legal obligations under POPCPI for children’s data and to culturally-grounded commitments to dignity-respecting design. Several noted that they maintained these protections even when they believed that parents favored continuous monitoring. Similarly, 
P12-PM mentioned that their product team implemented \textit{``higher privacy standards''} to protect children's data, deciding that data collected by smart baby monitors would only be streamed in real time and not stored. These decisions were presented as a negotiation between perceived user attitudes, relevant legal frameworks (e.g., POPCPI), and broader cultural expectations. However, P17-EN noted the absence of specific data protection measures for children, explaining that from an algorithmic standpoint, there is no clear distinction between adult and child data with respect to privacy considerations.

\textbf{Prioritizing UX requirements gathering.} Product teams, particularly product managers and UX/user researchers, employed various methods to collect requirements for the design and development of features and functionalities of smart home devices. 
They emphasized the importance of involving users to improve usability and functionality, as well as to identify practical issues during the initial requirements gathering and validation stages. 
In large product teams, some participants outsourced requirements gathering by asking third parties to conduct interviews and focus groups with users on their behalf; however, P5-UX criticized the industry’s reliance on convenience sampling, arguing that it limited research findings to individual perspectives shaped by narrow or biased participant pools: \textit{``The biggest challenge is that the people we interview can only represent themselves---they cannot speak for all users.''} 
Notably, privacy communication was managed by legal officers or product managers rather than UX/user researchers or marketing staff. Marketing participants (P25–P27) reported having limited input on privacy and security, largely because of minimal interaction with design and development teams. Even when probed, none reported involvement in communicating privacy features to users. Further, unlike typical software development, smart home device development often prioritizes hardware requirements (e.g., sensor integration)
. Some participants observed this shift, noting that it could result in privacy and security being de-prioritized. For example, P8-UX viewed privacy and security primarily as software issues rather than integral parts of the overall design process.

\textbf{Limited attention to privacy \& security in testing.} 
Many participants reported that they focused on assessing usability, stability, functionality, performance, and efficiency during the testing phase. Privacy and security were often neglected. 
In contrast, the teams of P9-UX, P10-PM, and P23-LE avoided involving external users to \textit{``protect trade secrets}''---relying instead on internal testing conducted by product team members. Only P18-LE mentioned regular collaboration with the testing team on data compliance before product launch, stating: \textit{``We are the final checkpoint for the product. It must pass through us before reaching the market, making compliance essential. I see our role not as setting the upper limit but as defining the minimum standard.''}

\textbf{Complying with applicable laws and regulations.}
Participants recognized the crucial role of legal compliance teams in shaping privacy practices. In addition to compliance experts, many participants, primarily product managers and senior leaders 
(e.g., software directors), noted that they frequently collaborated with legal compliance colleagues to assess security and privacy risks and ensure alignment with 
laws and regulations. 
As P12-PM noted: \textit{``Our legal team typically reviews the final designs for compliance before entering the development phase. They stay proactive by evaluating products 
whenever there are regulatory updates.''} 
Some participants mentioned that their legal teams focused primarily on reviewing device privacy policies but did not always verify whether privacy requirements were fully met. P14-EN also expressed concerns about complying with new laws or regulations: \textit{``I know some small companies might rush to launch smart home apps before new laws take effect, trying to dodge compliance by exploiting regulatory loopholes. They focus more on getting to market first than on adhering to the upcoming regulations.''} 

In addition, we observed that participants with over ten years of experience and in managerial roles (e.g., P5-UX, P15-EN, \& P24-EN) demonstrated greater awareness of privacy and data protection laws compared to less experienced participants. They advised their product teams to adjust feature designs to comply with legal requirements. For example, P5-UX altered the user consent dialog interface on their mobile app for smart home cameras to align with POPCPI regulations\footnote{Under POPCPI Articles 8 and 9, data processors are required to safeguard children’s personal data and obtain consent from their legal guardians prior to processing.}: \textit{``The law requires us to obtain consent from parents or guardians, so we implemented a dedicated consent option specifically for them. We also included a `refusal' option in case the guardian does not agree to provide authorization.''}

\textit{\textbf{Summary.}} Most participants equated gathering privacy \& security requirements with complying with Chinese privacy laws that prioritized national security over individual privacy; see \S\ref{subsec:regulations}. In contrast, gathering UX requirements---covering functionality and performance---involved active user engagement, often outsourced to third parties. Privacy was applied selectively and contextually---driven by (i) legal requirements (e.g., POPCPI for children’s data) and (ii) cultural or dignity values (e.g., respectful blurring in elder-care scenarios). Privacy and security received limited focus during testing.

\subsubsection{\emph{Personal} perspectives and practices} \label{Individual}

We also explored the \emph{personal} views and practices of our participants concerning multi-user smart home privacy and security.

\textbf{Strong commitment to privacy and ethics.} 
Most participants 
demonstrated a strong commitment to user privacy and ethics, with empathy that guided their decisions and helped them balance ethical considerations, technical implementation, and user privacy protection. 
For example, P1-EN noted: \textit{``If the company plans to develop a product that poses risks of infringing on user privacy, I will decline to participate in its development, even if it negatively affects my KPI.''} Furthermore, some shared that their own discomfort with being monitored by smart home devices directly influenced their approach to privacy-focused development. These personal experiences shaped their efforts to prioritize user privacy in the products they helped create. 
For example, 
P12-PM noted that: \textit{``informed user consent, transparent data practices, and voluntary data sharing are key components.''} 

\textbf{User ownership and control of data in multi-user smart homes.} Participants emphasized the importance of user ownership rights, particularly regarding access to, management of, and control over their data. By granting or revoking permissions, users could determine who had access to their data and when, helping the primary user maintain control over their smart home while balancing convenience with privacy and security. 
For example, some participants developing smart home cameras, doorbells, and door locks described how their design assigned a \textit{``super administrator''} role to a single account. 
P11-PM noted: \textit{``The owner can grant permissions to friends, domestic workers, or even strangers. These permissions may include activating certain features, such as viewing real-time video streams or accessing stored video clips, but only within predefined time periods.''} 

\textbf{Limited exposure to legal and privacy training.} Most participants reported some engagement with legal data privacy and protection training programs, although they noted that the training was often superficial. For instance, P14-EN from a small PE mentioned: \textit{``Our internal training focused more on real-life development cases rather than providing in-depth legal knowledge.''} Similarly, P6-UX expressed skepticism about the focus of these programs: \textit{``These training programs seem to be aimed at protecting the company, not the users.''} P13-UX also described the training as generic and not specifically tailored to address the privacy needs of smart home camera users: \textit{``We have an online exam on confidentiality or permissions every month, or every one to two months.''} 
Further, many participants relied heavily on their companies' legal or regulatory team members to handle compliance issues, often viewing legal compliance as separate from the main design, development, and testing phases (\S\ref{ProductTeams}). 
Interestingly, P18-LE highlighted their team's proactive role in delivering training sessions: \textit{``We often provide training for data compliance and privacy, security laws and regulations [...]. Some teams may invite us for smaller and more focused training sessions. All training sessions are recorded for future reference.''}

\textbf{Recommendations to improve privacy practices.} 
Many participants suggested that companies should increase transparency in data practices, ensuring users were clearly informed about data collection, sharing, and deletion. For example, 
P1-EN, P4-PM, and P17-EN advocated for an ``over-design'' strategy: \textit{``Designing privacy features involves adapting to evolving laws and user needs by going beyond standard compliance---implementing stricter rules, enhanced privacy controls and disclosures, and improved data handling practices.''} While P2-PM and P14-EN mentioned the need for a “top-down” approach to defining privacy requirements---such as establishing a clear development road-map---P14-EN stated: \textit{``It is essential to involve privacy regulation experts during the initial design and development phases, rather than only during or after testing.''} Additionally, some participants highlighted future development challenges related to integrating AI into smart home devices. For example, P14-EN, P17-EN, and P24-EN noted that embedding generative AI (e.g., Deepseek) in smart home devices could introduce new privacy concerns, as also reported in the academic literature~\cite{zhan2023privacy}. 

\textbf{Perceptions of Chinese users' privacy awareness and cultural influences.} While several participants recognized the importance of addressing privacy needs in China's smart home market, most believed that privacy awareness among Chinese users varied significantly, influenced by cultural and regional factors. Some felt that only a small fraction of the population was highly privacy-conscious, actively choosing products with strong privacy features, such as creating guest accounts or using surveillance devices only when away from home. Others thought that user privacy concerns were largely influenced by where they resided, with P25-MS stating: \textit{``Residents in more developed areas, such as Shanghai and other first-tier cities, tend to have greater privacy awareness compared to those in less developed regions.''} 
Several participants believed that most Chinese users were relatively unaware of or unconcerned about privacy risks compared to users in Western countries: \textit{``People with higher privacy awareness constitute even less than 1\%''} (P14-EN). Participants (e.g., P11-PM \& P14-EN) framed collectivism as one influence among several---such as professional ethics, personal values, and market demands---rather than as the sole driver of their privacy practices. 
P14-EN explained, \textit{``The education we receive from a young age emphasizes collectivism, and placing too much emphasis on personal interests is viewed as selfish in our cultural values.''} This cultural context led some Chinese product teams to downplay privacy in their product designs, assuming that Chinese users prioritized collective benefits over individual privacy. Furthermore, P19-LE discussed the influence of national laws on individual privacy awareness, stating: \textit{``Data compliance regulations in China differ from those in Europe. In China, national security takes precedence in data compliance, leading individuals to prioritize national security over their own data privacy and security.''} 

Many participants emphasized the necessity of collecting personally identifiable information (PII), such as phone numbers, for account registration to comply with China's real-name system under CSL. Since phone numbers are tied to real-world identities, participants believed that this helped reduce security risks by allowing law enforcement to trace malicious activity. Additionally, several noted that the widespread use of CCTV and facial recognition in public spaces across China had fostered general acceptance of such technologies, resulting in fewer concerns about their use in domestic settings. This contrasted with participants’ perceptions of Western contexts, where they believed that heightened privacy concerns could impede the adoption of home surveillance devices~\cite{wang2011concerned,li2022cultural}.

\textit{\textbf{Summary.}} Participants' personal experiences with smart homes and the discomfort of being monitored had nurtured empathy for users and the desire to balance privacy with functionality. However, many believed that Chinese users prioritized collective benefits over individual privacy, leading them to de-prioritize privacy in product design. Yet, privacy was negotiated between perceived user attitudes, legal requirements, and socio-cultural norms, with some participants prioritizing it on ethical grounds even when they believed users would not.

\subsection{Development Challenges and Power Dynamics} \label{challenges}

\subsubsection{Development challenges} \label{challenges in development}
We outline the challenges participants and their teams faced, including limited attention to at-risk groups, difficulty balancing privacy with UX, and broader constraints such as limited resources and development pressures. Some participants also distinguished between data essential to core device functions (e.g., camera footage for incident detection or remote monitoring, access-control logs) and non-essential data (e.g., personalization signals and analytics). For essential data, they emphasized necessity, proportionality, and minimal retention, whereas for non-essential data, they favored opt-in defaults, granular controls, and data minimization. 

\textbf{Challenges in designing for at-risk groups.}
Participants faced significant challenges in gathering requirements from at-risk groups such as children, older adults, and domestic workers. Limited resources prevented participants from conducting direct research with these users, making it difficult to understand and address their specific needs. For example, P17-EN noted that their research prioritized ``\emph{the majority population in China},'' often excluding groups such as survivors of intimate partner violence (IPV). P5-UX and P7-PM pointed out the challenges of recruiting IPV survivors and gaining their trust, as many were reluctant to disclose personal experiences. This 
made it especially difficult to design effective and inclusive solutions for vulnerable populations.
Furthermore, P6-UX described the challenges their team faced when designing smart TVs for older and disabled users, citing a lack of experience in understanding the specific privacy concerns of these groups. P4-PM's team implemented protective measures for older adults: \textit{``We use millimeter-wave radars in our cameras that capture only coarse-grained data, such as blurred body shapes, to protect privacy, especially in elderly care homes.''}

\textbf{Challenges in balancing privacy and UX.} Although most participants acknowledged the importance of privacy, they also recognized the necessity of collecting data to enhance products, services, convenience, and user experience. This created a development paradox: emphasizing privacy required collecting less user data, which, in turn, could restrict improvements to UX. As P24-EN said: \textit{``Collecting extensive user data is essential for improving our products; without it, providing personalized and optimized UX becomes more difficult.''} Some participants mentioned that many Chinese users were willing to trade privacy for convenience, posing a significant challenge for product teams aiming to balance the two. For example, P9-UX noted that long privacy policies could undermine usability, while P11-PM protected user privacy/security through robust measures: \textit{``For security reasons, we implemented complex password requirements to access cameras. However, we received considerable negative feedback in the app store, and many users found the process cumbersome and not user-friendly.''}

\textbf{Resource constraints.} Resource limitations also pushed some companies to depend on third-party services for cloud storage and speech recognition, making it harder to maintain control over data and privacy practices. 
Additionally, some participants acknowledged the use of dark patterns and consent manipulation in the design of smart home devices. P9-UX expressed concerns about certain \textit{``rogue companies''} that concealed or omitted key information within lengthy privacy policies: 
\textit{``Users must click on blue links within privacy policies and follow multiple redirects to find specific info about the identity and location of third-party services used.''} 
To tackle these challenges, P6-UX suggested 
using visual aids: \textit{``We can shorten lengthy, dull privacy statements by highlighting key points in different colors to improve clarity and understanding.''} Interestingly, P4-PM admitted that their product team deliberately concealed the in-app account deletion option to prevent users from accidentally deleting their baby monitor accounts---an example of a common dark pattern practice~\cite{bosch2016tales}: \textit{``We concealed the account deletion option within the app, requiring users to contact customer service if they wanted to delete their accounts.''}

\subsubsection{Power Dynamics}\label{power dynamic} We examined participant's perceptions of power dynamics in the Chinese smart home ecosystem---between companies, product teams, authorities, and users---and explored power imbalances at the household level, particularly those shaped by factors such as gender. 

\textbf{Companies vs. companies.}
Some participants highlighted the complex and often challenging relationships between small and large tech companies, particularly when integrating into tech giants' ecosystems, which imposed strict compliance and quality standards. 
P10-PM and P12-PM elaborated on these challenges: 
\textit{``Joining a tech giant's smart home ecosystem forces small firms to meet strict standards that may conflict with user preferences and hinder integration. Without specialized expertise, these firms often face compatibility issues and must invest extra effort to align with larger platforms.''} P18-LE argued that integration into tech giants’ ecosystems often compelled smaller firms to comply with strict standards, not only because of regulatory requirements but also out of concern for maintaining their reputation: \textit{``Compliance from these small companies is essential not only for meeting legal standards, but also for maintaining the reputation of larger players.''}

\textbf{Companies vs. product teams.} Many participants emphasized that the dynamics of internal power within their companies often restricted the autonomy of product teams, making it difficult to prioritize or implement privacy measures. Senior management, marketing, or business units frequently had greater influence, leading to decisions that favored marketability or speed over user privacy. 
For instance, P15-EN noted: \textit{``This decision-making process often compelled us to prioritize company objectives over user-centered design and privacy-focused innovations.''} Further, some participants indicated that internal policies and corporate culture played a major role in 
how privacy measures were implemented. 

\textbf{Users vs. users.} Many participants pointed out the complex power dynamics present in multi-user smart homes, especially tensions between parental monitoring and children's privacy, hosts and guests, as well as employers and domestic workers. For example, P3-PM, P12-PM, and P26-MS described conflicts emerging from the use of smart home cameras in parent-teen relationships, conflicts shaped by Confucian values stressing filial piety and deference to parental authority, as P3-PM explained: \textit{``Many Chinese parents buy these cameras primarily to help with homework and monitor their children's posture while studying [...]. Children often find it difficult to challenge the decisions of their parents [...] 
In Chinese families, questioning the choice of a parent is often seen as disrespectful.''} 
In addition, within Chinese culture, where hospitality is highly valued, guests may either be unaware that they are being monitored or feel uneasy knowing that smart devices are present, which could be perceived as a breach of trust and courtesy.

Additionally, P9-UX, P18-LE, and P20-LE noted the growing use of smart home devices to monitor domestic workers 
for security or deterrence purposes, raising ethical concerns about their privacy: \textit{``Privacy and security settings are typically designed to protect device owners, often overlooking the comfort of domestic workers who lack the power to consent to being monitored by their employers' surveillance cameras.''} Further, P12-PM observed that purchasing and installing smart home devices often mirrored existing household gender roles, stating
: \textit{``Gender dynamics often shape buying decisions in typical Chinese households: When a woman suggests buying a device, she may seek input from other family members, but if a man, usually the father or husband, disagrees, the purchase often does not happen.''}

\textbf{Users vs. companies.} Several participants highlighted the difficulties smart home users encountered in protecting their privacy, noting that companies frequently prioritized business interests over user privacy, especially within China's existing regulatory environment. As a result, many users felt helpless when their privacy was violated, particularly if they lacked legal expertise or resources. For example, P5-UX pointed out the challenges users faced in gathering evidence after a data breach involving smart home devices (e.g., due to lack of resources or limited knowledge). 
P19-LE added: \textit{``Under current laws, companies are rarely held liable unless there is clear economic harm. Additionally, authorities tend to avoid arbitrarily imposing penalties, since frequent fines could disrupt business operations and reduce tax revenues.''}

\textbf{Law enforcement vs. companies.} Some participants noted that their companies usually required legal documents, such as a search warrant, to authorize police access to user data, in accordance with PIPL (Articles 34, 35). P5-UX added that they generally avoided granting access to data unless it was essential for an investigation, sometimes obtaining the consent of the homeowner beforehand. 
However, some participants noted that certain companies allowed police access to all user registration and facial data collected by smart home cameras in specific emergency situations. For example, P13-UX highlighted: \textit{``Government or public security departments have special privileges to access user facial data, names, and other personal information. They have the authority to force us to provide this data and can also retrieve it through internal channels, bypassing the restrictions our teams put in place.''} 

\textit{\textbf{Summary.}} Participants faced several challenges: limited resources for designing for at-risk groups in mind, tensions between data collection and privacy, and navigating complex power dynamics in the Chinese smart home ecosystem. Gender roles and family hierarchies---rooted in filial piety---often discouraged younger members from questioning why older members monitored them using smart devices.
\subsection{
Interventions}
\label{tab:mitigating}


\subsubsection{Technical interventions} \label{Technical}

To mitigate privacy concerns and comply with laws and regulations, many participants described anonymizing or blurring image data prior to processing. 
As P9-UX explained, \textit{``We don't capture real images directly---sensitive data, like nude bodies, is processed by blurring privacy zones detected by smart cameras.''} 
Furthermore, several company guidelines imposed strict procedures for managing data, especially when it involved sensitive items such as biometric records. For example, P11-PM and P24-EN noted that camera footage was encrypted prior to transmission, and their team built specialized decoders to retrieve those recordings. P9-UX and P17-EN described employing a data-minimization approach, with P9-UX explaining: \textit{``We only send and save posture data, like simplified 2D images resembling stick figures, instead of actual scene details.''} Moreover, many participants mentioned that their companies followed data deletion practices, often in accordance with regulatory mandates. For example, P4-PM, P16-PM, and P19-LE noted that Chinese regulatory bodies required companies to respect users' rights to have their data erased. P5-UX added: \textit{``Given the large volume of data and the high cost of storage, we typically delete video footage after it has been streamed in real time, unless users have subscribed to a storage service.''} P11-PM, P21-EN, and P22-EN also mentioned that users had the option to delete their cloud-stored data within 30 days, after which the system would automatically remove it.

Most participants highlighted that their companies imposed strict controls on sensitive user information, allowing only authorized personnel to access it. For example, P12-PM explained: \textit{``Our software developers have different levels of access. For example, a developer with a user account can view linked devices, whereas database and server developers are limited to operational queries, such as tracking user numbers or feature usage, without accessing individual user data. Once a feature is deployed, it is controlled by the user, and software developers no longer need to maintain or access it.''} 

However, some participants, especially UX designers, expressed a lack of clarity regarding how smart home devices managed and processed data, often citing limited access and strict company policies as the reason. As P6-UX stated: \textit{``We often lack a full understanding of data collection and processing because our design team operates separately from the compliance team.''} P18-LE and P23-LE also underscored this issue from the compliance team's perspective, noting limited cross-team communication. 

\subsubsection{Suggested social interventions} \label{Social}

Many participants stressed the need for collaboration between smart home companies, government agencies, academic institutions, and privacy experts to create effective strategies and standards aimed at improving user privacy awareness and protection. For example, P10-PM highlighted the value of academic involvement in connecting industry practices with research insights, stating: \textit{``Universities and companies could hold annual meetings to discuss and integrate academic insights into industry practices.''} However, participants admitted that merging academic expertise with industry practices can be challenging, as P15-EN noted: \textit{``Certain experts don't fully grasp the practical realities of industry, which makes the standards they develop feel useless in practice. 
It often feels like ``non-experts'' are driving the development.''} 

In addition, some participants emphasized the importance of better privacy education for the general public in China, believing that many Chinese users tend to view privacy rights as less important for ordinary people than for celebrities or the wealthy. Several participants recommended that the government launch community outreach initiatives to raise privacy awareness and encourage best practices among users. For example, P14-EN suggested: \textit{``If possible, local sub-district communities could hold privacy protection workshops that require the participation of all community members.''} P6-UX additionally highlighted the importance of SOEs in leading privacy education initiatives: \textit{``Major SOEs should educate the public on privacy, data usage, and protection strategies. We could include privacy handbooks alongside smart home device manuals and distribute these materials to device owners and purchasers.''} Further, P10-PM emphasized the need to integrate privacy education into the Nine-Year Compulsory Education curriculum\footnote{~\href{http://en.moe.gov.cn/documents/laws_policies/201506/t20150626_191391.html}{Compulsory Education Law of the People's Republic of China}}, stating that doing so would foster awareness of privacy from a young age and equip future generations with the knowledge to better understand and safeguard their privacy rights. 

\subsubsection{Suggested legal interventions} \label{Legal}
Many participants highlighted the need for clearer legal regulations around data management. For example, P10-PM pointed out that vague legal terminology made it difficult to determine when privacy breaches occurred. P1-EN suggested: \textit{``Clearly defined legislation with stipulated penalties for privacy violations could prompt companies to adopt a more cautious approach during development.''} 
In contrast, P5-UX, P12-PM, and P19-LE argued that laws should be intentionally broad to encompass a wider array of innovative practices, ensuring they stayed effective as new data collection methods emerged. A small number of participants shared their experiences designing smart homes for overseas customers. For example, P14-EN noted: \textit{``Privacy laws in China are not as comprehensive as GDPR. We should work on adapting international standards to fit the local context. Although Chinese privacy regulations have been revised to keep pace with technological and social changes, they often fall behind, resulting in gaps and challenges in enforcement.''} P1-EN, P2-PM, and P27-MS talked about China's strict regulatory environment, which restricted foreign services and closely monitored the privacy practices of domestic companies, to protect national security. 

Additionally, most participants highlighted the central government's crucial role in overseeing data privacy and security, emphasizing its significant impact on how companies handled privacy. As P9-UX remarked, \textit{``Organizations at the community level, such as 'juweihui', are essential to raise awareness about privacy and security, helping individuals to better grasp applicable regulations.''} Some participants highlighted the crucial role national organizations could play in protecting user data by establishing privacy standards. For example, P1-EN suggested: \textit{``Certain national standards bodies could establish a multi-level system for data management, classifying devices into low-, medium-, or high-risk categories, with each category subject to distinct compliance and verification requirements.''}

\textit{\textbf{Summary.}} Participants used technical measures to comply with Chinese privacy laws and regulations, but acknowledged a shortage of adequate social and legal interventions, leading some to suggest improvements in these areas.

%% file: content/5-discussion.tex
\section{Discussion}\label{discussion}
\subsection{Key Findings}
Participants showed a strong commitment to legal compliance and ethical design in the early design and development stages, regularly holding review meetings to ensure compliance with Chinese privacy and data protection laws, treating this as part of privacy-related requirements gathering. However, in the final testing phases, usability and functionality were prioritized over privacy. Privacy–convenience trade-offs also emerged when teams collected data that was not essential to core device functions (e.g., for personalization), where UX was sometimes prioritized over privacy. Many participants, especially UX designers often
had little to no formal training in legal compliance, suggesting limited understanding of privacy laws and regulations, whereas most product managers displayed greater knowledge of legal compliance and data practices. Some participants' personal experiences with smart home devices helped them develop empathy for users. However, their practices were heavily shaped by Chinese cultural values, which tend to minimize individual privacy in favor of convenience and collective benefits \textbf{(RQ1)}.

Our participants encountered three main challenges: 1) low public awareness of privacy issues affecting at-risk groups, 2) finding a balance between privacy and UX, and 3) managing power dynamics within the Chinese multi-user smart home ecosystem. These challenges were especially significant in multi-user settings, where coordinating consent and privacy among multiple stakeholders was complex. We also found that Chinese cultural norms, shaped by Confucian values such as filial piety, limited children's ability to contest parental monitoring through smart home devices out of respect for parental authority. Furthermore, some companies granted emergency police access to all user registration and facial data, and government or public security agencies could forcibly obtain this information, potentially circumventing company-imposed restrictions \textbf{(RQ2)}.

Our participants employed various privacy mitigation strategies, including technical measures like restricting data access internally and adopting decentralized data handling. They also identified gaps between current legal frameworks and real-world development challenges, prompting calls for clearer regulations to improve smart home privacy and security. Additionally, they recommended several social and legal interventions \textbf{(RQ3)}.

\subsection{Contributions} \label{contributions}

Although some of our findings align with existing research showing that software developers often prioritize functionality and usability over security and privacy~\cite{nurgalieva2023narrative,acar2017developers,tahaei2020understanding,tahaei2021privacy,tahaei2021developers,chowdhury2021developers,tahaei2022charting,tahaei2023stuck,acar2016you,gutfleisch2022does,klemmer2023make}, we specifically examine the challenges faced by product teams in the context of smart home design and development. This setting introduces distinct privacy concerns due to its intimate environment and complex multi-user interactions. In addition, our participants---who held diverse product team roles beyond software development---navigated additional complexities, including balancing the needs of different stakeholder groups, managing hierarchical family relationships, and operating within an evolving and often ambiguous regulatory framework in a non-WEIRD context. In some cases, their practices---shaped by local cultural and familial norms---extended beyond complying with legal requirements, underscoring how regulatory frameworks, cultural values, and contextual factors jointly influenced team decision-making (\S\ref{power dynamic}). For example, whereas Western research typically suggests individually tailored privacy controls to safeguard each household member's autonomy, our findings show participants created paternalistic features allowing parents to remotely monitor their children through `family supervision devices' (e.g., child desk lamps). This design approach reflects Chinese cultural values such as filial piety and collective well-being, differing from Western norms where privacy preferences are usually managed independently by each household member without a central family authority controlling surveillance. Although our study does not provide a direct comparison between the privacy practices of Western and non-Western smart home product teams, it highlights culturally-embedded design choices within Chinese smart home teams that may contrast with dominant assumptions in Western-centric privacy literature.

Compared to prior studies on smart home product design and development outside China (e.g., \cite{chalhoub2020innovation,chalhoub2024useful,albayaydh2023examining,albayaydh2024co}), our research uniquely examines how Chinese smart home product teams navigate their distinct regulatory and socio-cultural environment (\S\ref{tab:views and practices} and \S\ref{challenges}). Whereas Albayaydh and Flechais \cite{albayaydh2024innovative} emphasized localization and contextual norms, our study shifts the focus inward to company practices, analyzing Chinese product teams and tracing how privacy is operationalized through requirements gathering, design trade-offs, and the prioritization of compliance with Chinese privacy laws. Moreover, our findings do not only shed light on the Chinese context but also offer insights relevant to other non-WEIRD regions facing similar cultural and regulatory dynamics. For example, in Southeast and South Asia, culturally adaptive privacy-by-design frameworks can help align product development with local norms and evolving standards~\cite{ha2023digital,sambasivan2018privacy} (\S\ref{consider_other_market}). Also, we enrich the literature by providing in-depth empirical evidence about the privacy perspectives/practices of product team members in diverse roles and organizations, including start-ups and state-owned enterprises, offering a comprehensive view of the smart home industry. Although earlier Chinese privacy research has focused on \emph{end-user} privacy perspectives~\cite{huang2019perception,liu2022privacy}, our study distinctively highlights the points of view of \emph{product team members}, illustrating how they manage privacy within their cultural context (\S\ref{Individual}, \S\ref{power dynamic}).

Furthermore, China’s extensive government oversight of data access imposes significant compliance burdens, often compelling product teams to provide data to state authorities~\cite{li2024brussels,wang2024justifying}. Building on this prior work, our study offers a comprehensive perspective on how product teams navigate these constraints and trade-offs in their daily design and development practices. By distinguishing between essential and non-essential data, we show how Chinese product teams’ decisions produced different types of trade-offs: essential data often raised questions of minimization and proportionality, whereas non-essential data heightened tensions between added convenience and user privacy. We also found that Chinese small and medium-sized companies typically relied on legal teams’ ecosystem-level checklists, rather than adopting systematic privacy engineering or threat modeling practices~
\cite{deng2011privacy,tahaei2022understanding} (see \S\ref{power dynamic}). The widespread acceptance of state surveillance in China reduces the emphasis on comprehensive privacy threat modeling, highlighting a culturally specific norm that remains underexplored in Western-centric research. 

\subsection{Chinese Social, Cultural, and Legal Factors}
\label{chinesecontext}

\subsubsection{Influence by Collectivism and Confucianism} Collectivist values could, in principle, motivate a shared responsibility among product team members to protect the privacy of end-users (as well as bystanders)~\cite{brewer2007collectives}. Yet, most of our participants did not frame their roles in these terms; instead, they typically aligned their privacy practices with legal compliance or market demands. Thus, our findings do not necessarily suggest that privacy is unimportant in China, or that Chinese smart home device end-users prioritize national security over individual privacy; rather, they illustrate how product teams’ perceptions of Chinese end-users, laws, and cultural norms shape the way privacy is operationalized in practice. 
Further, some participants felt that ordinary people’s privacy mattered less than that of the wealthy---a view rooted in Confucian values, prioritizing hierarchy and social roles, where privacy is often considered less critical for lower-status individuals~\cite{hofstede1984culture,hofstede2011dimensionalizing,sambasivan2018privacy,wang2015confucianism,wei2013confucian}.


While the notion of ``yin si'' (privacy) is less central in Chinese culture compared to Western liberal thought, Confucian values---especially those emphasizing family duties---influenced the decisions of smart home product team members. Teams reflected these values by designing features that promoted family cohesion and enabled parental or elder care (\S\ref{power dynamic}). For example, they added functions allowing children to remotely monitor and support their aging parents, aligning with filial piety norms and parental respect~\cite{ho1996filial}. 
Devices also included features for parental monitoring, such as tracking study progress. Participants noted that users generally viewed such monitoring as a helpful parenting tool, not as invasive. This aligns with Confucian values that tie academic achievement to filial duty~\cite{naftali2010caged,wu1996parental,liu2024cctv,shi2023monkey}, and where respect for hierarchy may discourage youth from resisting such monitoring practices~\cite{wang2022individual}. These choices contrast with Western design norms, which prioritize individual privacy and autonomy, including stricter protections for children's data and requirements like explicit consent, reflecting a stronger focus on individual rights over collective family control.

To address power imbalances in Chinese smart homes and tensions between Confucian values and privacy, participants prioritized individual privacy in specific cases, guided by virtues like empathy and compassion~\cite{slote2010mandate}. They implemented protections for children's PII to meet regulatory demands (\S\ref{ProductTeams}). Unlike Western norms that broadly center individual privacy, Chinese teams balanced legal requirements with cultural values. While supporting household-wide monitoring aligned with collectivism, they added privacy safeguards for vulnerable users when required. This reflects a nuanced approach in which Confucian ideals of family harmony coexist with legal and ethical pressures, shaping distinct design practices.

\subsubsection{Challenges due to normalized surveillance}
Based on \S\ref{Individual}, participants noted that widespread CCTV use in China has normalized surveillance, reducing public sensitivity to monitoring~\cite{su2022explains}. Framed by the government as necessary for safety and order, constant exposure has led to greater acceptance of data collection. Over time, this may shift societal norms, making surveillance routine and lowering expectations for personal privacy and data security~\cite{chin2022surveillance,chen2023maintainers}. 
This could also influence how people perceive surveillance in private spaces, with some accepting public and private monitoring as normal, while others expect more privacy in their homes.

\subsubsection{Community-based education}
To raise privacy awareness, we highlight the importance of community-based agencies and sub-districts as grassroots platforms (\S\ref{Legal}). Backed by Residents Committees (RCs, juweihui), which play a key role in community management and education, these platforms can reach hundreds to thousands of residents~\cite{wang2018democratic}. We recommend broad community training programs and privacy protection lectures 
for all community members
; and distributing privacy-focused materials such as handbooks and posters. Although primarily aimed at the general public, product team members, being part of the broader population, can also benefit from these resources. Furthermore, product teams would benefit from formal company-led training or workshops (\S\ref{training suggestion}).

\subsubsection{Clear and enforceable privacy laws} Participants (\S\ref{ProductTeams}, \S\ref{power dynamic}, \S\ref{Legal}) noted that applying foreign legal standards in the context of Chinese smart home product design is challenging due to vague privacy regulations, unclear data handling protocols, and the difficulty of proving privacy breaches in court~\cite{cui2021legal}. These gaps leave users vulnerable, with little legal recourse and high burdens of proof. Although China's PIPL outlines clearer definitions for concepts like consent and data portability~\cite{yao2023overcoming}, its enforcement is largely left to individual companies. This has led to inconsistent compliance practices, with firms often relying on internal interpretations and informal guidance rather than standardized protocols. Government-corporate power dynamics, particularly regarding state access to user data, further complicate efforts to balance regulatory compliance, business responsibilities, and individual privacy. 
To respond, participants advocated for clearer legal definitions and practical implementation frameworks aligned with China's regulatory context and industry conditions (e.g., government-approved third-party enforcement).
A tiered compliance structure, customized to company size, sector, and risk level, would offer a pragmatic path toward enforceable and trusted privacy protections in the Chinese smart home industry.


\subsection{Improving Privacy Practices of Product Teams}
\label{privacyindevelopement}
\subsubsection{Implement clear privacy guidelines} Participants highlighted difficulties in embedding strong privacy and security features into smart home devices, driven by rising user expectations for transparent data practices in China (\S\ref{challenges in development} and \S\ref{Individual}), echoing trends observed in WEIRD contexts~\cite{alomar2022developers,ekambaranathan2020understanding,chalhoub2020innovation,chalhoub2024useful}. Although national standards such as GB/T 35273-2020 prohibit default privacy policy consent checkboxes~\cite{gb35723}
, many UX designers in our study struggled with implementing appropriate data handling and avoiding privacy-invasive elements like dark patterns~\cite{bosch2016tales,gray2018dark}, making it harder to communicate privacy terms clearly. To address this, we recommend creating detailed, practical design guidelines that incorporate privacy-by-design principles~\cite{cavoukian2009privacy,hoepman2014privacy,schaub2015design} and emphasize privacy from the early stages of development, especially in multi-user environments. These resources should feature real-world use cases and expert input (e.g., privacy champions~\cite{tahaei2021privacy}) to support designing with privacy as a core principle, fostering a culture where privacy is built-in rather than added later. 

\subsubsection{Enhance privacy compliance in small businesses} We found that most large companies---including SOEs, PEs, and MNCs---generally followed privacy regulations during product development (\S\ref{ProductTeams}, \S\ref{power dynamic}, and \S\ref{Legal}), restricting team access to sensitive user data and enforcing internal privacy policies (\S\ref{Individual}). Large SOEs, under stricter government scrutiny and public accountability, adhered to more rigorous standards. While large PEs showed more flexibility, they still complied with national laws. In contrast, smaller businesses often prioritized business needs over privacy and legal compliance due to limited resources, echoing findings in prior research~\cite{kekulluouglu2023we,tahaei2022embedding,albayaydh2024innovative}. 
To support smaller companies, we suggest that Chinese regulators develop clear and enforceable frameworks aligned with domestic privacy laws and national standards (e.g., PIPL -- Chapter 6). These should include specific guidance for small businesses, clarify enforcement authority, offer support resources, propose strategies for managing limited capacities, and introduce incentives for compliance. Establishing community-based agencies and industry associations can further promote adherence to legal requirements.


\subsubsection{Consider power dynamics and at-risk groups} Based on findings from \S\ref{ProductTeams} and \S\ref{power dynamic}, we observed limited awareness among product teams regarding the privacy needs of at-risk groups in China. Men in Chinese households frequently held decision-making power and device ownership in smart homes, positioning women in the same household as `passenger user'~\cite{kraemer2020further,geeng2019s,strengers2019protection,despres2024my,albayaydh2023examining}, echoing broader patriarchal norms. Teams also struggled to engage with at-risk populations like domestic workers~\cite{bernd2022balancing,slupska2022they,albayaydh2022exploring,he2025exploring} due to limited access, ethical complexities, and workers' reluctance to share information~\cite{sannon2022privacy}. These barriers hinder the creation of inclusive smart home technologies. To address this, we recommend ethical, user-centered design guidelines that reflect the privacy needs of all household members, especially women and vulnerable users. Government-led efforts (e.g., CAC and Women's Federation) and local bodies such as RCs can offer targeted support~\cite{he2025living}. We also call on researchers to build frameworks that systematize and translate at-risk users' concerns into actionable insights for product design.

\subsubsection{Encourage privacy training and collaboration} \label{training suggestion} Participants prioritized meeting legal needs, but often relied heavily on specialized compliance staff, as noted in \S\ref{Individual}. This over-reliance led to limited legal awareness among broader product teams regarding their responsibilities. While many participants expressed empathy for users' privacy needs, this alone was insufficient without clear internal guidelines and a solid grasp of legal frameworks. To address this, we recommend collaborative work among team members early in product design, supported by strong corporate policies that guide teams through the evolving privacy challenges in Chinese multi-user smart homes. Companies should also develop structured training initiatives---such as \emph{mandatory} lectures/courses that address legal compliance and ethical concerns rooted in cultural contexts~\cite{knapp2009information}. These should be part of on-boarding learning and ongoing professional development, featuring real-life case studies that present ethical trade-offs and cultural tensions around user privacy, helping teams turn empathy into actionable strategies by aligning legal knowledge with ethical judgment. %
We also encourage partnerships between companies, academia, and policymakers to co-develop standardized educational programs that highlight both global best practices and local cultural nuances. 

\subsubsection{Consider privacy compliance in other markets} \label{consider_other_market} As discussed in \S\ref{Legal}, participants emphasized the need for a deeper understanding of legal and cultural differences across international markets (e.g., EU countries) in order to design smart home devices that are both compliant and aligned with local privacy norms. We recommend adopting a user-centered, culturally-adaptive privacy-by-design framework or approach that includes conducting privacy impact assessments with culturally-specific adaptations, which would be beneficial in contexts or countries where collectivist values and weaker individual privacy regulations dominate. Engaging users from different regions---through collaboration with local stakeholders and culturally diverse usability testing---can help identify privacy and usability concerns. Customizing privacy settings for specific markets is especially important in regions where privacy regulations are still developing, such as Southeast Asia, the Middle East, and Africa~\cite{albayaydh2024innovative,albayaydh2024co,sambasivan2018privacy,ha2023digital,mienye2024artificial}.

Our findings also revealed that a small number of our participants (e.g., P12) were involved in designing smart home products for overseas customers. These participants developed stronger privacy awareness through engagement with international regulations such as GDPR. This exposure shaped their global compliance strategies and influenced domestic design practices, with GDPR often serving as a benchmark for meeting or exceeding standards in other regions, such as India or Latin America. However, this transnational orientation created tensions between international legal expectations and local user norms or marketability concerns. To navigate these challenges, participants emphasized the value of internal workshops to help teams better understand and localize privacy expectations across regions. 

\subsection{Future Work}
Future research should investigate privacy and security issues from the perspectives of at-risk individuals in Chinese multi-user smart home environments. Participants from Chinese smart home product teams anticipated the emergence of AI-driven features---such as real-time voice recognition and generative LLM-based dialogue in smart speakers and cameras~\cite{deepseek2025}. While these advancements may enhance UX, they also raise significant concerns regarding passive data collection and user profiling~\cite{zhan2023privacy,zhan2024healthcare}. Future work should further examine smart home privacy policies and design practices, with particular attention to the widening gaps between these practices and Chinese privacy laws and regulations (e.g., PIPL).

%% file: content/6-conclusion.tex
\section{Conclusion}
Our interviews with 27 members of Chinese smart home product teams provide new insights into their privacy perspectives and practices in a non-WEIRD context. Participants demonstrated a strong commitment to legal compliance and ethical considerations within China’s strict legal and regulatory environment. However, they also faced challenges in navigating the complexities of an evolving regulatory and technological landscape, including limited understanding of the privacy needs of at-risk groups, gender disparities, and power imbalances among authorities, companies, and users. Moreover, our findings highlight how cultural and social norms in China shape perceptions of privacy. Smart home product teams often perceived privacy as a lower priority for users---an attitude influenced by the collectivist orientation of Chinese society---which might have contributed to the de-prioritization of privacy in product design and development.

We recommend the creation of clear guidelines for transparent data practices in smart home devices, emphasizing the protection of end-users’ privacy, particularly for those at heightened risk. Achieving this requires balancing individual privacy rights with national security considerations. In addition, we call for privacy-focused training and educational initiatives for both users and smart home product teams in China, who may operate under the assumption that the general public places limited value on privacy.